\documentclass[twocolumn,PRA]{revtex4}
\usepackage{graphicx}

\def\duzomniejsze{<\kern-.7mm<}
\def\duzowieksze{>\kern-.7mm>}

\def\textbf#1{{\bf #1}}
\def\beq{\begin{equation}}
\def\eeq{\end{equation}}
\def\be{\begin{equation}}
\def\ee{\end{equation}}
\def\ben{\begin{eqnarray}}
\def\een{\end{eqnarray}}
\def\beqa{\begin{eqnarray}}
\def\eeqa{\end{eqnarray}}
\def\eea{\end{array}}
\def\bea{\begin{array}}
\newcommand{\bei}{\begin{itemize}}
\newcommand{\eei}{\end{itemize}}
\newcommand{\bee}{\begin{enumerate}}
\newcommand{\eee}{\end{enumerate}}

\def\>{\rangle}
\def\<{\langle}

\begin{document}

\title{Nonclassicality of quantum excitation of classical coherent field in photon loss channel}

\begin{abstract}
{\normalsize We investigate the nonclassicality of photon-added
coherent states in the photon loss channel by exploring the
entanglement potential and negative Wigner distribution. The total
negative probability defined by the absolute value of the integral
of the Wigner function over the negative distribution region
reduces with the increase of decay time. The total negative
probability and the entanglement potential of pure photon-added
coherent states exhibit the similar dependence on the beam
intensity. The reduce of the total negative probability is
consistent with the behavior of entanglement potential for the
dissipative single-photon-added coherent state at short decay
times.

}

\end{abstract}

\author{Shang-Bin Li$^{2}$}\email{stephenli74@yahoo.com.cn}, \author{Xu-Bo
Zou$^{1}$}, \author{Guang-Can Guo$^{1}$}

\affiliation{1. Key Laboratory of Quantum Information, University
of Science and Technology of China, Hefei 230026, China. }

\affiliation{2. Shanghai research center of Amertron-global,
Zhangjiang High-Tech Park, \\
299 Lane, Bisheng Road, No. 3, Suite 202, Shanghai, 201204, P.R.
China}

\maketitle

Nonclassical optical fields play a crucial role in understanding
fundamentals of quantum physics and have many applications in
quantum information processing \cite{Bouwmeester}. Usually, the
nonclassicality manifests itself in specific properties of quantum
statistics, such as the antibunching \cite{Kimble}, sub-poissonian
photon statistics \cite{Short}, squeezing in one of the
quadratures of the field \cite{Dodonov}, partial negative Wigner
distribution \cite{Hillery}, etc..

When the nonclassical optical fields propagate in the medium, they
inevitably interact with their surrounding environment, which
causes the dissipation or dephasing \cite{Gardiner}. It is well
known that the dissipation or dephasing will deteriorate the
degree of nonclassicality of the optical fields. A quantitative
measure of non-classicality of quantum fields is necessary for
further investigating the dynamical behavior of their
non-classicality. Many authors have investigated the relations
between non-classicality of optical fields and the entanglement
and shown that non-classicality is a necessary condition for
generating inseparable state via the beam splitter \cite{Kim2002}.
Based on them, a measure called the entanglement potential for
quantifying the non-classicality of the single-mode optical field
has been proposed \cite{Asboth2005}. The entanglement potential is
defined as the entanglement achieved by 50:50 beam splitter
characterized by the unitary operation
$U_{BS}=e^{\frac{\pi}{4}i(a^{\dagger}b+ab^{\dagger})}$ acting on
the target optical mode $a$ and the vacuum mode $b$. Throughout
this paper, log-negativity is explored as the measure of
entanglement potential. The log-negativity of a density matrix
$\rho$ is defined by \cite{vidalwerner} \be
N(\rho)=\log_2\|\rho^{\Gamma}\|, \ee where $\rho^{\Gamma}$ is the
partial transpose of $\rho$ and $\|\rho^{\Gamma}\|$ denotes the
trace norm of $\rho^{\Gamma}$, which is the sum of the singular
values of $\rho^{\Gamma}$.

Nevertheless, experimental measurement of the entanglement
potential is still a challenge task. How to quantify the variation
of the nonclassicality of quantum fields based on the current
mature laboratory technique is an interesting topic. The Wigner
function is a quasi-probability distribution, which fully
describes the state of a quantum system in phase space. The
partial negativity of the Wigner function is indeed a good
indication of the highly nonclassical character of the state.
Reconstructions of the Wigner functions in experiments with
quantum tomography \cite{Vogel,Smithey,Welsch} have demonstrated
appearance of their negative values, which can not be explained in
the framework of the probability theory and have not any classical
counterparts. Therefore, to seek certain possible monotonic
relation between the partial negativity of the Wigner distribution
and the entanglement potential may be an available first step for
experimentally quantifying the variation of nonclassicality of
quantum optical fields in dissipative or dephasing environments.

Here, for clarifying the feasibility of this idea, we investigate
the nonclassicality of photon-added coherent states in the photon
loss channel by exploring the entanglement potential and negative
Wigner distribution. The total negative probability defined by the
absolute value of the integral of the Wigner function over the
negative distribution region is introduced, and our calculations
show it reduces with the increase of decay time. The total
negative probability and the entanglement potential of pure
photon-added coherent states exhibit the similar dependence on the
beam intensity. The reduce of total negative probability is
consistent with the behavior of entanglement potential for the
dissipative single-photon-added coherent state at short decay
times.

The photon-added coherent state was introduced by Agarwal and Tara
\cite{Agarwal1991}. The single photon-added coherent state (SPACS)
is experimentally prepared by Zavatta et al. and its nonclassical
properties are detected by homodyne tomography technology
\cite{Zavatta2004}. Such a state represents the intermediate
non-Gaussian state between quantum Fock state and classical
coherent state (with well-defined amplitude and phase)
\cite{Glauber}. For the SPACS, a quantum to classical transition
has been explicitly demonstrated by ultrafast time-domain quantum
homodyne tomography technique. Thus, it is timely to analyze how
the photon loss affects the non-classicality of such kind of
optical fields.

Let us first briefly recall the definition of the photon-added
coherent states (PACSs) \cite{Agarwal1991}. The PACSs are defined
by $\frac{1}{\sqrt{N(\alpha,m)}}a^{\dagger{m}}|\alpha\rangle$,
where $|\alpha\rangle$ is the coherent state with the amplitude
$\alpha$ and $a^{\dagger}$ is the creation operator of the optical
mode. $N(\alpha,m)=m!L_m(-|\alpha|^2)$, where $L_m(x)$ is the
$m$th-order Laguerre polynomial. When the PACS evolves in the
photon loss channel, the evolution of the density matrix can be
described by \cite{Gardiner} \be
\frac{d\rho}{dt}=\frac{\gamma}{2}(2a\rho{a}^{\dagger}-a^{\dagger}a\rho-\rho{a}^{\dagger}a),
\ee where $\gamma$ represents dissipative coefficient. The
corresponding non-unitary time evolution density matrix can be
obtained as \beqa
\rho(t)&=&\frac{1}{m!L_m(-|\alpha|^2)}\sum^{\infty}_{k=0}\frac{(1-e^{-\gamma{t}})^k}{k!}\nonumber\\
&&\hat{L}(t)a^ka^{\dagger{m}}|\alpha\rangle\langle\alpha|a^ma^{\dagger{k}}\hat{L}(t),\eeqa
where $\hat{L}(t)=e^{-\frac{1}{2}\gamma{t}a^{\dagger}a}$. For the
dissipative photon-added coherent state in Eq.(3), the total
output state passing through a 50/50 beam splitter characterized
by the unitary operation
$e^{\frac{\pi}{4}i(a^{\dagger}b+ab^{\dagger})}$ with a vacuum mode
b can be obtained or \beqa
\rho_{tot}=D_a(t)D_b(t)\frac{e^{-m\gamma_2{t}}e^{|\alpha|^2(e^{-\gamma_2{t}}-1)}}{m!L_m(-|\alpha|^2)}\sum^{\infty}_{k=0}\frac{(e^{\gamma_2{t}}-1)^k}{k!}\nonumber\\
\hat{E}^k\hat{E}^{\dagger{m}}|00\rangle\langle00|\hat{E}^m\hat{E}^{\dagger{k}}D^{\dagger}_a(t)D^{\dagger}_b(t),\eeqa
where
$D_a(t)=e^{\frac{\sqrt{2}}{2}\alpha(t){a}^{\dagger}-\frac{\sqrt{2}}{2}\alpha^{\ast}(t){a}}$
and
$D_b(t)=e^{\frac{\sqrt{2}i}{2}\alpha(t){b}^{\dagger}+\frac{\sqrt{2}i}{2}\alpha^{\ast}(t){b}}$
are the displacement operators of the modes a and b, respectively,
where $\alpha(t)=\alpha{e}^{-\gamma_2{t}/2}$.
$\hat{E}=\frac{\sqrt{2}}{2}a-\frac{\sqrt{2}i}{2}b+\alpha{e}^{-\frac{1}{2}\gamma_2{t}}$.
The local unitary operators can not change entanglement,
therefore, we only need to consider the entanglement of the mixed
state given as follows: \beqa
\rho^{\prime}_{tot}=\frac{e^{-m\gamma_2{t}}e^{|\alpha|^2(e^{-\gamma_2{t}}-1)}}{m!L_m(-|\alpha|^2)}\sum^{\infty}_{k=0}\frac{(e^{\gamma_2{t}}-1)^k}{k!}\nonumber\\
\hat{E}^k\hat{E}^{\dagger{m}}|00\rangle\langle00|\hat{E}^m\hat{E}^{\dagger{k}}.\eeqa
The log-negativity of the above density matrix can be analytically
solved for the case of single photon excitation, i.e. $m=1$, but
its expression is still lengthy. The corresponding entanglement
potential of the SPACSs quantified by log-negativity is given in
Fig.5(b).

On the other hand, the presence of negativity in the Wigner
function of the field is also the indicator of nonclassicality.
The Wigner function, the Fourier transformation of characteristics
function \cite{Barnett} of the state can be derived by
\cite{Cessa} \be
W(\beta)=\frac{2}{\pi}{\mathrm{Tr}}[(\hat{O}_e-\hat{O}_o)\hat{D}(\beta)\rho\hat{D}^{\dagger}(\beta)],\ee
where $\hat{O}_e\equiv\sum^{\infty}_{n=0}|2n\rangle\langle2n|$ and
$\hat{O}_o\equiv\sum^{\infty}_{n=0}|2n+1\rangle\langle2n+1|$ are
the even and odd parity operators respectively. In the photon loss
channel described by the master equation (2), the time evolution
Wigner function satisfies the following Fokker-Planck equation
\cite{Carmichael} \beqa
\frac{\partial}{\partial{t}}W(q,p,t)&=&\frac{\gamma}{2}(\frac{\partial}{\partial{q}}q+\frac{\partial}{\partial{p}}p)W(q,p,t)\nonumber\\
&&+\frac{\gamma}{8}(\frac{\partial^2}{\partial{q}^2}+\frac{\partial^2}{\partial{p}^2})W(q,p,t).\eeqa
where $q$ and $p$ represent the real part and imaginary part of
$\beta$, respectively. Substituting the initial Wigner function of
a SPACS \cite{Agarwal1991} \be
W(q,p,0)=\frac{-2L_1(|2q+2ip-\alpha|^2)}{\pi{L_1}(-|\alpha|^2)}e^{-2|q+ip-\alpha|^2}\ee
and the initial Wigner function of a two photon-added coherent
state (TPACS) \cite{Agarwal1991} \be
W(q,p,0)=\frac{2L_2(|2q+2ip-\alpha|^2)}{\pi{L_2}(-|\alpha|^2)}e^{-2|q+ip-\alpha|^2}\ee
into the Eq.(7), we can obtain the time evolution Wigner function.
In Fig.1, the Wigner function of the SPACS with $\alpha=0.5$ at
three different values of decay time are plotted. The phase space
Wigner distribution of the pure SPACS with $\alpha=0.5$ loses its
circular symmetry and moves away from the origin because of the
appearance of a definite phase. The partial negativity of the
Wigner function indicates the nonclassical nature of the single
quantum excitation of the classical coherent field. The photon
loss causes the gradual disappearance of the partial negativity of
the Wigner function. The tilted ringlike wings in the distribution
gradually start to disappear and the distribution becomes more and
more similar to the Gaussian typical of a classical coherent
field. In Fig.2, the phase space Wigner distributions at $p=0$ of
the SPACSs in the photon loss channel are depicted for several
different values of $\alpha$ and $\gamma{t}$, from which the
influence of photon loss on the partial negativity of the Wigner
function is explicitly shown. For the cases of
$\alpha=0.1,0.5,1.0,1.5$, it is found that those curves $W(q,0)$
at decay times $\gamma{t}=0,0.2,0.4,0.6$ exhibit the partial
negativity. The further photon loss will completely destroy the
partial negativity. One can also find that, the larger the
parameter $|\alpha|$, the more rapidly the Wigner function tends
to the Gaussian function.
\begin{figure}
\centerline{\includegraphics[width=2.5in]{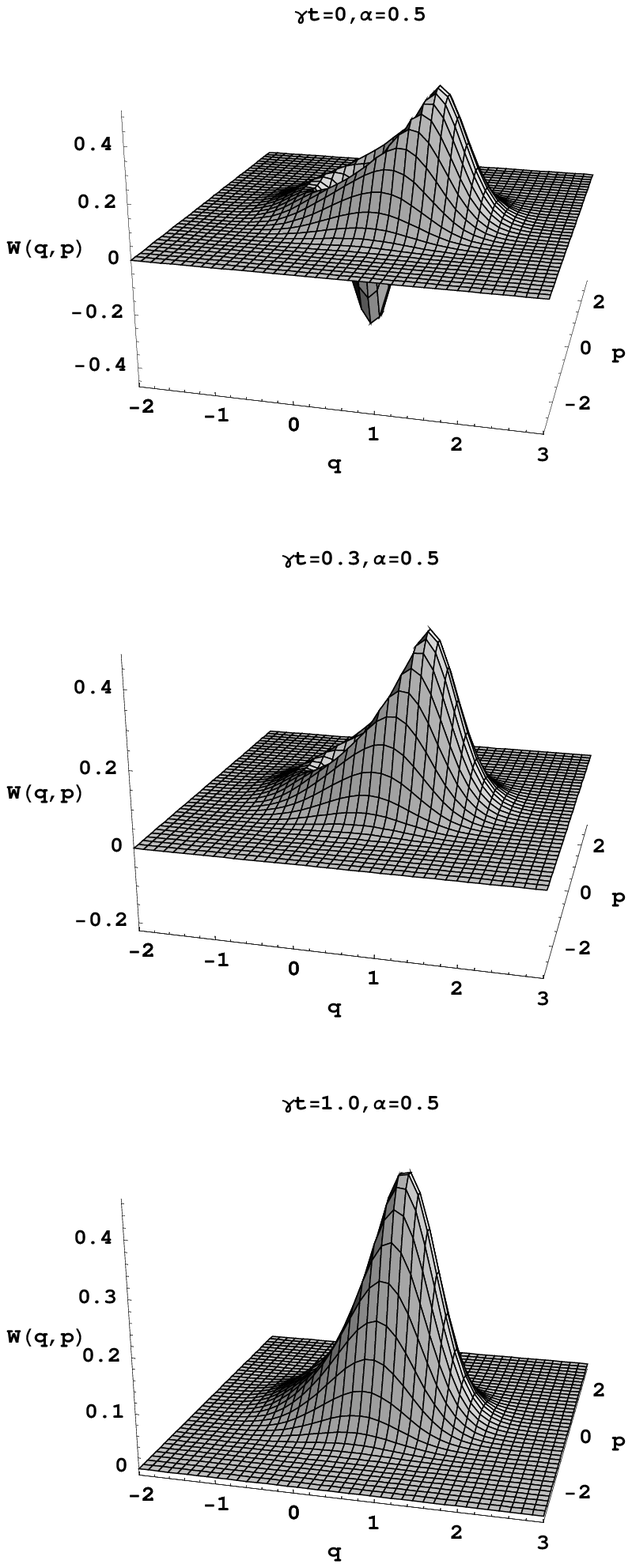}}
\caption{The Wigner functions of the SPACS with $\alpha=0.5$ in
photon loss channel are depicted for three different values of
decay time $\gamma{t}$.}
\end{figure}
\begin{figure}
\centerline{\includegraphics[width=2.5in]{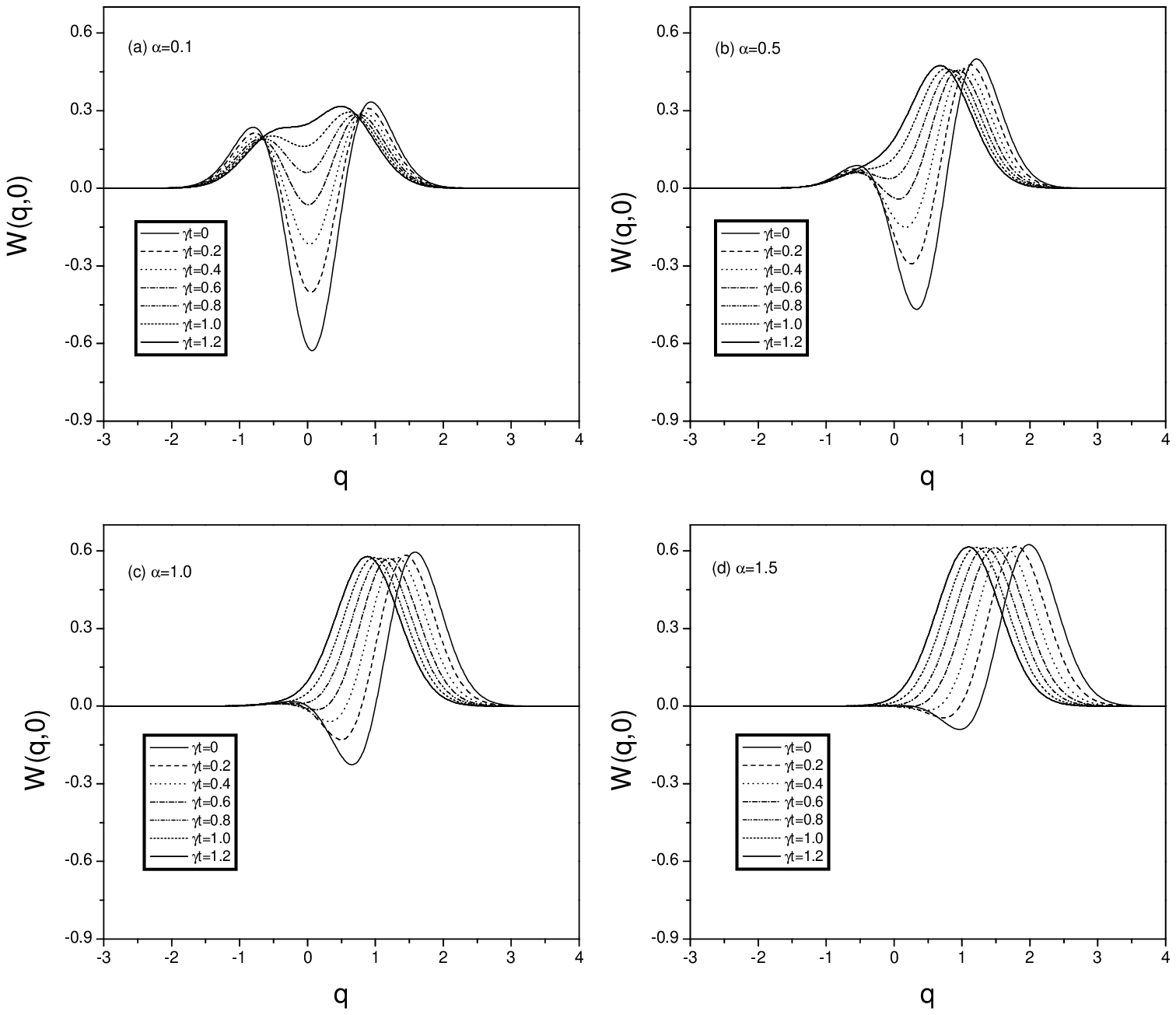}}
\caption{The Wigner distribution function at $p=0$ of the SPACSs
with four different values of initial amplitudes $\alpha$ in the
photon-loss channel are depicted for several decay times
$\gamma{t}$. From bottom to top, the decay times $\gamma{t}$ are
0, 0.2, 0.4, 0.6, 0.8, 1, 1.2, respectively. (a) $\alpha=0.1$; (b)
$\alpha=0.5$; (c) $\alpha=1.0$; (d) $\alpha=1.5$.}
\end{figure}
\begin{figure}
\centerline{\includegraphics[width=2.5in]{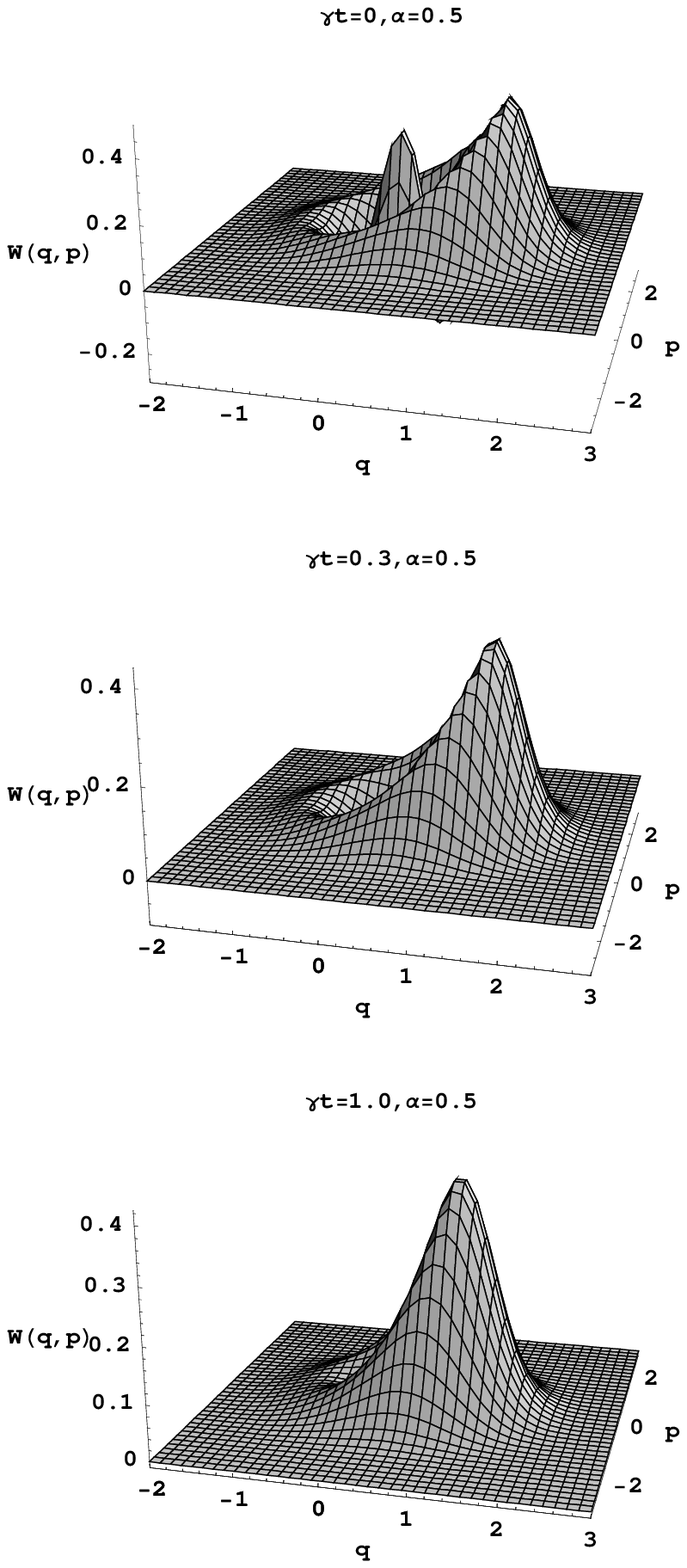}}
\caption{The Wigner functions of the TPACS with $\alpha=0.5$ in
photon loss channel are depicted for three different values of
decay time $\gamma{t}$.}
\end{figure}
\begin{figure}
\centerline{\includegraphics[width=2.5in]{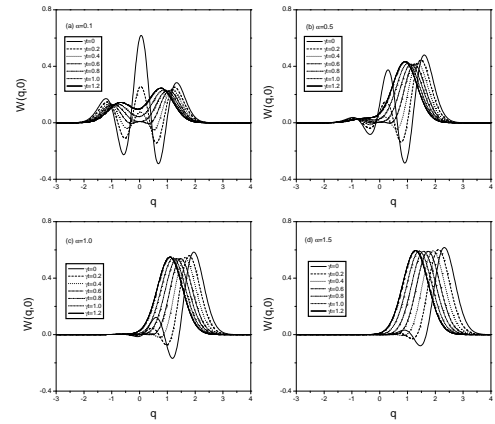}}
\caption{The Wigner distribution function at $p=0$ of the TPACSs
with four different values of initial amplitudes $\alpha$ in the
photon-loss channel are depicted for several decay times
$\gamma{t}$. From bottom to top, the decay times $\gamma{t}$ are
0, 0.2, 0.4, 0.6, 0.8, 1, 1.2, respectively. (a) $\alpha=0.1$; (b)
$\alpha=0.5$; (c) $\alpha=1.0$; (d) $\alpha=1.5$.}
\end{figure}
The pure TPACSs exhibit more non-classicality than the pure SPACSs
when the entanglement potential is adopted as the measure of
nonclassicality \cite{Devi2006}. In Fig.3, the Wigner functions of
TPACSs with the parameter $\alpha=0.5$ at three different values
of decay times are depicted. It is also shown that the photon loss
deteriorates its partial negativity. For more explicitly observing
the details, we plot $W(q,0)$ of the TPACSs with different values
of $\alpha$ and $\gamma{t}$ in Fig.4. Different from the cases of
SPACSs, here, $W(q,0)$ of these pure TPACSs with $|\alpha|\leq1$
have two explicit negative local minimal values. As $|\alpha|$
increases, the absolute value of the negative local minimum at the
left more rapidly decreases than the one at the right. As
$\gamma{t}$ increases, the absolute value of negative minimum of
$W(q,0)$ decreases which implies the decreases of the
non-classicality of the states. When $\gamma{t}$ exceeds a
threshold value, the partial negativity of the Wigner distribution
can not be explicitly observed from this figure. Similarly, the
larger the parameter $|\alpha|$, the more rapidly the Wigner
function of the TPACSs in the photon-loss channel tends to the
Gaussian function.
\begin{figure}
\centerline{\includegraphics[width=2.5in]{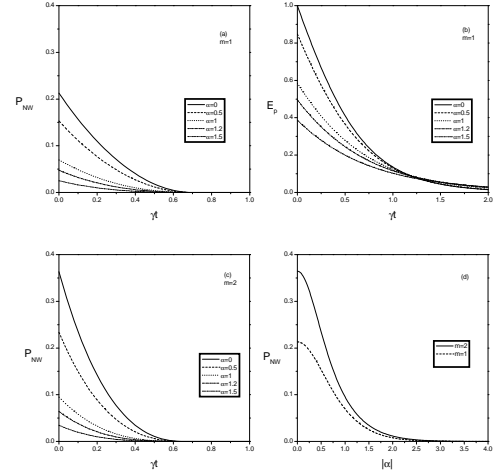}}
\caption{(a) The absolute values $P_{NW}$ of negative Wigner
distribution probability of the SPACSs with several different
values of initial amplitude $\alpha$ are plotted as the function
of decay time $\gamma{t}$ in photon loss channel. (b) The
entanglement potential measured by log-negativity of the SPACSs in
photon-loss channel are plotted as the function of $\gamma{t}$.
(c) The absolute values $P_{NW}$ of negative Wigner distribution
probability of the TPACSs with several different values of initial
amplitude $\alpha$ are plotted as the function of decay time
$\gamma{t}$ in photon loss channel. (d) $P_{NW}$ of the pure
SPACSs and the pure TPACSs are plotted as the function of
$|\alpha|$. }
\end{figure}
A natural question arises whether the partial negativity of the
Wigner function can be used to quantitatively measure the
non-classicality of certain kinds of nonclassical fields. It is
obvious that the negative minimum of the Wigner function can not
appropriately quantitatively measure the nonclassicality. For
example, comparing Fig.2(a,b) with Fig.4(a,b) respectively, the
negative minimum in the Wigner function of the pure SPACSs is
smaller than the one of the pure TPACSs. However, there was
already the evidence that the pure TPACSs possesses larger
nonclassicality than the pure SPACS with the same $|\alpha|$
\cite{Devi2006}. Nevertheless, the absolute value $P_{NW}$ of the
total negative probability of the Wigner function may be a good
choice for quantifying the nonclassicality. $P_{NW}$ is defined by
\be P_{NW}=|\int_{\Omega}W(q,p)dqdp|,\ee where $\Omega$ is the
negative Wigner distribution region, and $|x|$ represents the
absolute value of $x$. In Fig.5(a), we plot $P_{NW}$ of the SPACSs
with different values of $\alpha$ as the function of $\gamma{t}$.
$P_{NW}$ decreases with $\gamma{t}$, and becomes invisible after a
threshold value of $\gamma{t}\simeq0.7$. In Fig.5(b), the
entanglement potential quantified by log-negativity of SPACSs in
the photon-loss channel is plotted. The entanglement potential
also decreases with $\gamma{t}$. These calculations partially
elucidate the consistent between $P_{NW}$ and the entanglement
potential of the dissipative SPACSs at short time. Currently, the
experimental quantitative investigation of the nonclassicality of
the quantum optical fields is still an open issue, and the
experimental measurement of entanglement potential has still
several technical difficulties. So, the measurement of $P_{NW}$
may be adopted as a replaced approach to investigate the influence
of photon loss on the nonclassicality of the SPACSs. In Fig.5(c),
we also investigate the effect of photon loss on $P_{NW}$ of the
TPACSs with various values of $\alpha$. At short decay times
$\gamma{t}\ll1$, the values of $P_{NW}$ of the TPACSs are larger
than those of the SPACSs with the same beam intensity
$|\alpha|^2$. However, $P_{NW}$ of the TPACSs is more fragile
against photon loss than the one of the SPACS. As an illustration,
for $\alpha=0.1$ and $\gamma{t}\geq0.34$, the $P_{NW}$ of TPACSs
is smaller than the one of SPACS.

From Fig.5(b), we can also find that the entanglement potentials
of SPACSs with different values of $|\alpha|<1$ more rapidly
decrease than the ones of SPACSs with $|\alpha|>1$ in the photon
loss channel. Both the pure SPACSs and pure TPACSs are
non-gaussian nonclassical states with partial negativity of the
Wigner distributions for any large but finite values of
$|\alpha|$. In Fig.5(d), the dependence of $P_{NW}$ of the pure
SPACSs and the pure TPACSs on $|\alpha|$ are shown. It is found
that both the $P_{NW}$ of the pure SPACSs and the pure TPACSs
reduce with the increase of $|\alpha|$, and the TPACS possesses
larger $P_{NW}$ than the SPACS with the same value of $|\alpha|$.
This property is also coincident with the dependence of
entanglement potentials of the pure SPACSs and the pure TPACSs on
$|\alpha|$ \cite{Devi2006}.

In summary, we have investigated the non-classicality of photon
excitation of classical coherent field in the photon loss channel
by exploring both the entanglement potential and the partial
negativity of the Wigner function. The total negative probability
defined by the absolute value of the integral of the Wigner
function over the negative distribution region is introduced.
Consistent behaviors of the total negative probability and the
entanglement potential of single quantum excitation of classical
coherent fields are found for the short time photon loss process.
Similar dependence of the total negative probability and the
entanglement potential on the beam intensity is revealed for few
photon-added coherent states. The partial negativity of the Wigner
function can not be observed when the decay time exceeds a
threshold value, while the entanglement potential always exists
for any large but finite decay time. In the future tasks, it is
interesting to investigate the variation of nonclassicality of
general nonclassical states in gaussian channel or non-gaussian
channel.

\section * {ACKNOWLEDGMENTS}

This work was supported by National Fundamental Research Program,
also by National Natural Science Foundation of China (Grant No.
10674128 and 60121503) and the Innovation Funds and
\textquotedblleft Hundreds of Talents\textquotedblright\ program
of Chinese Academy of Sciences and Doctor Foundation of Education
Ministry of China (Grant No. 20060358043)

\end{document}